%% file: main.tex
\documentclass[runningheads]{llncs}
\usepackage[utf8]{inputenc}
\usepackage[T1]{fontenc}
\usepackage{graphicx} 
\graphicspath{ {./images/} }

\input{addon/pkgloading}%
\input{addon/squeeze}
\input{addon/glossary}

\input{addon/glossary-BPM}

\input{addon/macros}

\begin{document}

\title{CAKE: Sharing Slices of Confidential Data on Blockchain}
\titlerunning{CAKE}

\author{%
    Edoardo~Marangone\inst{1}\orcidlink{0000-0002-0565-9168}
	\and
    Michele~Spina\inst{1}\orcidlink{0009-0003-6870-5525}
	\and
    Claudio~Di~Ciccio\inst{2}\orcidlink{0000-0001-5570-0475}
	\and
    Ingo~Weber\inst{3}\orcidlink{0000-0002-4833-5921}
}
\authorrunning{Marangone et al.}

\institute{%
    Sapienza University of Rome, Rome, Italy\\
	\email{\href{mailto:marangone@di.uniroma1.it}{marangone@di.uniroma1.it}};
	\email{\href{mailto:spina.1711821@studenti.uniroma1.it}{spina.1711821@studenti.uniroma1.it}}
	\and
    Utrecht University, Utrecht, Netherlands\\
	\email{\href{mailto:c.diciccio@uu.nl}{c.diciccio@uu.nl}}
	\and
	Technical University of Munich, School of CIT, and Fraunhofer Gesellschaft, Munich, Germany\\
	\email{\href{mailto:ingo.weber@tum.de}{ingo.weber@tum.de}}
}

\maketitle

\begin{abstract}
	\input{sections/abstract}
\end{abstract}

\section{Introduction}
\label{sec:intro}
\input{sections/intro}

\section{Core concepts and tool architecture}
\label{sec:architecture}
\input{sections/architecture}

\section{Demonstration through a real-world case study}
\label{sec:demostrating}
\input{sections/demostrating}

\section{Implementation}
\label{sec:implementation}
\input{sections/implementation}

\section{Related work}
\label{sec:sota}
\input{sections/sota}

\section{Conclusion and future developments}
\label{sec:conclusion}
\input{sections/conclusion}

\smallskip
\footnotesize{\mysubsubsection{Acknowledgements.}
	The work of E.~Marangone and C.~Di~Ciccio was funded by projects 
	PINPOINT (B87G22000450001), under the PRIN MUR program, and BRIE (Cyber 4.0).
}
\vspace{-1\baselineskip}

\bibliographystyle{splncs04}
\bibliography{bibliography}

\end{document}

%% file: addon/pkgloading.tex
\usepackage[english]{babel}
\usepackage{subcaption}
\usepackage{url}
\usepackage{colortbl}
\usepackage{amssymb}
\usepackage{stmaryrd}
\usepackage{amsmath}
\usepackage{mathrsfs}
\usepackage{amssymb}
\usepackage[left]{lineno}
\usepackage{xfrac}
\usepackage{nicefrac}
\usepackage[textsize=scriptsize,backgroundcolor=yellow!40,obeyFinal]{todonotes}
\usepackage[nonumberlist,acronym,sanitize=none]{glossaries}
\glsdisablehyper
\usepackage{comment}
\usepackage[pdftex, colorlinks=true, hyperfootnotes=true, hyperindex=true,
            plainpages=false, pagebackref=false, pdfpagelabels=true, pdfstartview=FitH,
            linkcolor=blue, citecolor=blue, urlcolor=blue,
            bookmarks, bookmarksopen, bookmarksdepth=3]{hyperref}
\usepackage[capitalise,nameinlink]{cleveref}
\crefname{algocf}{alg.}{algs.}
\Crefname{algocf}{Algorithm}{Algorithms}
\crefname{section}{Sect.}{Sects.}
\Crefname{section}{Section}{Sections}
\usepackage{tkz-base}
\usetikzlibrary{decorations.pathmorphing,trees,snakes,arrows,shapes,automata,petri}
\usepackage{paralist}
\usepackage{multicol}
\usepackage{booktabs}
\usepackage{enumitem}
\usepackage{multirow}
\usepackage{rotating}
\usepackage{etaremune}
\usepackage{marginnote}
\usepackage{mathtools}
\usepackage{mathabx}
\usepackage{adjustbox}
\usepackage{ifthen}
\usepackage[normalem]{ulem}
\usepackage{lineno}
\usepackage{soul}
\usepackage{floatrow}
\usepackage{listings}
\crefname{lstlisting}{Listing}{Listings}
\usepackage{lipsum}
\usepackage{makecell}
\usepackage{diagbox}
\usepackage[scientific-notation=false]{siunitx}
\usepackage{footmisc}
\usepackage{xspace}
\usepackage{ifdraft}
\usepackage{wrapfig}
\usepackage{orcidlink}

%% file: addon/squeeze.tex
\usepackage{newtxtext}

\usepackage[subtle]{savetrees}

%% file: addon/glossary.tex
\newacronym[\glslongpluralkey={Distributed Ledger Technologies}]{dlt}{DLT}{Distributed Ledger Technology}

\newacronym{ipfs}{IPFS}{InterPlanetary File System}

\newacronym{p2p}{P2P}{peer-to-peer}

\newacronym{abe}{ABE}{Attribute-Based Encryption}

\newacronym{maabe}{MA-ABE}{Multi-Authority Attribute-Based Encryption}

\newacronym{cpabe}{CP-ABE}{Ciphertext-Policy Attribute-Based Encryption}

\newacronym{ssl}{SSL}{Secure Sockets Layer}

\newacronym{dht}{DHT}{Distributed Hash Table}

\newacronym{mpc}{MPC}{Multi-party Computation}

\newglossaryentry{box}{ %
  name={Box},
  description={authority intialisation storage box},
  first={\glsentrydesc{box} (henceforth, \glsentrytext{Box} for short)},
  plural={Boxes},
  firstplural={\glsentrydesc{box}es (henceforth, \glsentryplural{box} for short)}
}

\newglossaryentry{rloc}{ %
	name={resource locator},
	description={a generic term to denote content-based links obtained via hashing (e.g., the IPFS link)},
}

\newacronym{cpmaabe}{CP-MA-ABE}{Ciphertext-Policy Multi-Authority Attribute-Based Encryption}

\newacronym{dk}{\textsl{dk}}{decryption key}
\newacronym{fdk}{\textsl{fdk}}{final decryption key}

%% file: addon/glossary-BPM.tex
\newacronym[\glslongpluralkey={Business Processes}]{bp}{BP}{Business Process}
\newacronym{bpi}{BPI}{Business Process Intelligence}
\newacronym{bpm}{BPM}{Business Process Management}
\newacronym{bpms}{BPMS}{Business Process Management System}
\newacronym{bpmn}{BPMN}{Business Process Model and Notation}
\newacronym{cpn}{CPN}{colored Petri net}
\newacronym{kpi}{KPI}{Key Performance Indicator}
\newacronym{ocbc}{OCBC}{Object-centric Behavioral Constraints}
\newacronym{soa}{SOA}{Service-Oriented Architecture}
\newacronym{pn}{PN}{Petri net}
\newacronym{wf}{WF}{workflow}
\newacronym{wfms}{WfMS}{Workflow Management System}
\newacronym{xes}{XES}{eXtensible Event Stream}
\newacronym{yawl}{YAWL}{Yet Another Workflow Language}
\newglossaryentry{task}{%
	name={task},description={the non-divisible, elementary activity}}

\newglossaryentry{promod}{%
	name={process model},description={the model of a process}
}
\def\LogAlph {\ensuremath{\Sigma}}
\newglossaryentry{logalph}{
	name={log alphabet},description={the process alphabet, as reflected in a log},%
	symbol={\LogAlph}}
\def\Evt {\ensuremath{e}}
\newglossaryentry{evt}{
	name={event},description={a record of an instantaneous fact during the process enactment},%
	symbol={\Evt}}
\def\Trc { \ensuremath{\tau} }

\newglossaryentry{trace}{
	name={trace},description={a sequence of \glsplural{evt}},%
	symbol={\Trc}}
\def\EvtLog {\ensuremath{L}}

\newglossaryentry{evtlog}{
	name={event log},description={a collection of \glstext{evttrace}s},%
	symbol={\EvtLog}}

%% file: addon/macros.tex
\renewcommand\tabcolsep{5pt}
\newcolumntype{d}{>{\columncolor{gray!10}}c}
\newcolumntype{m}{>{\columncolor{gray!10}}l}
\newfloatcommand{capbtabbox}{table}[][\FBwidth]
\newenvironment{iiilist}%
{\begin{inparaenum}[\itshape(i)\upshape]}%
{\end{inparaenum}}

\newcommand{\mysubsubsection}[1]{\noindent\textbf{#1}}

\RequirePackage{xparse}
\NewDocumentEnvironment{AuthNote}{+o+o}{%
	\IfValueT{#2}{\marginnote{\scriptsize{#2}}}%
	\begin{scriptsize}
		\colorbox{gray}%
		{\color{white} Note\IfValueT{#1}{ (#1)}:}%
		\quad%
		\color{brown}
}{%
	\normalcolor
	\end{scriptsize}
}

\RequirePackage{pifont}

\RequirePackage{lipsum}
\newcommand{\LipsumGray}[1][]{{\color{gray}\ifthenelse{\equal{#1}{}}{\lipsum}{\lipsum[#1]}}}

\RequirePackage{siunitx}
\newcolumntype{D}[1]{S[
	table-omit-exponent,
	round-mode=places,
	round-integer-to-decimal,
	round-precision={#1}]} 



%% file: sections/abstract.tex
Cooperative information systems typically involve various entities in a collaborative process within a distributed environment.
Blockchain technology offers a mechanism for automating such processes, even when only partial trust exists among participants.
The data stored on the blockchain is replicated across all nodes in the network, ensuring accessibility to all participants.
While this aspect facilitates traceability, integrity, and persistence, it poses challenges for adopting public blockchains in enterprise settings due to confidentiality issues. 
In this paper, we present a software tool named Control Access via Key Encryption (CAKE), designed to ensure data confidentiality in scenarios involving public blockchains.
After outlining its core components and functionalities, we showcase the application of CAKE in the context of a real-world cyber-security project within the logistics domain.

%% file: sections/intro.tex
Blockchain technology is increasingly being applied in information systems of diverse enterprise domains due to its capacity to facilitate the establishment and execution of cooperative processes involving multiple parties with limited mutual trust~\cite{Weber.etal/BPM2016:UntrustedBusinessProcessMonitoringandExecutionUsingBlockchain,Stiehle22SLR}.
The decentralized structure of public permissionless blockchains ensures that each participant in the network possesses a replicated ledger, thereby allowing for unrestricted accessibility of all data. This transparency, in conjunction with the immutability of data and the non-repudiable nature of transactions, makes blockchains a robust foundation for verifiable and trustworthy interactions.

In scenarios where there is a lack of mutual trust among parties, hiding some data from the majority of users can be advantageous. Indeed, when blockchain technology is discussed, the security and privacy topics are the critical issues and their importance is underlined and well recognized~\cite{Privacy1,Privacy2,Corradini.etal/ACMTMIS2022:EngineeringChoreographyBlockchain}.
A solution to guarantee data secrecy and confidentiality among parties was presented in~\cite{Marangone.etal/BPM2022:CAKE} under the name of Control Access via Key Encryption (CAKE). The parties can securely exchange information using the CAKE architecture, hiding data or parts thereof from others.
This paper demonstrates the CAKE tool, illustrating its implementation and the newly introduced features.
We used CAKE as a core component of a larger platform designed and realized in the context of a national cyber-security research and innovation project for international logistics: Blockchain Register for Import-Export (BRIE).%
\footnote{\label{foot:brie}\href{https://brie.moveax.it/en}{https://brie.moveax.it/en}, accessed 2024-03-11}
We employ the case study to showcase the maturity and integration of the tool within a real-world setting.
At large, our research provides security-minded practitioners with a tool to securely transact confidential data: A whole public blockchain network permanently stores the transactions attesting to the validity and integrity of the data, but only authorized parties can read the actual information in-clear.

In the following, \cref{sec:architecture} outlines the CAKE architecture and the core concepts it builds upon. In \cref{sec:demostrating}, we demonstrate our proof-of-concept implementation with the BRIE real-world use case. \Cref{sec:implementation} provides implementation details about our tool. \Cref{sec:sota} presents the related work in the literature. Finally, \cref{sec:conclusion} concludes the paper and draws some avenues for future work.

%% file: sections/architecture.tex
In the following, we outline the key methodologies and techniques underpinning our solution. Equipped with these notions, we describe the core components of CAKE thereafter.

\mysubsubsection{Core concepts.}
Distributed Ledger Technologies (DLTs) are protocols that facilitate transactional storage, processing, and validation within a decentralized network without the need for central authorities or intermediaries. These transactions come along with cryptographic signatures. The resulting shared transaction log collectively constitutes a ledger accessible to all participants in the network. In a \textbf{blockchain}, a specific type of DLT, transactions are organized in blocks, which are linked to form an append-only singly linked list, namely a chain. DLTs, including blockchains, are tamper-resistant thanks to cryptographic techniques such as hashing and decentralized validation of transactions. Public blockchain platforms such as 
Ethereum~\cite{Wood/2018:Ethereum} and Algorand~\cite{Chen.Micali/TCS2019:Algorand} require the payment of fees for submitting and processing transactions on the platform. These platforms enable the utilization of \textbf{smart contracts}, which are programs deployed, stored, and executed directly on-chain~\cite{Dannen/2017:IntroducingEthereumandSolidity,zheng2020overview}. Ethereum and Algorand support smart contracts through the Ethereum Virtual Machine (EVM) and the Algorand Virtual Machine (AVM), respectively. These contracts are deployed and invoked via transactions. Their code is stored on the blockchain and executed by the nodes within the distributed system. The results of contract invocations are subject to blockchain consensus, thereby being verified by the network and completely traceable.
To reduce the costs associated with invoking smart contracts, external Peer-to-Peer (P2P) systems are employed for storing significant volumes of data~\cite{Xu.etal/2019:ArchitectureforBlockchainApplications}. Among the facilitating technologies is the \textbf{InterPlanetary File System (IPFS)}.
IPFS is a distributed system utilizing a Distributed Hash Table (DHT) to distribute stored files across multiple nodes. 
It employs hashing to generate a uniquely identifying resource locator for every file. 
In a conventional blockchain integration, the locator is subsequently transmitted to a smart contract for permanent storage on the blockchain~\cite{Lopez-Pintado.etal/IS2022:ControlledFlexibilityBlockchainCollaborativeProcesses}. Notice that such an address is content-based: changing even a single bit in the data entails the modification of the hash, thus the original locator does not match the modified data.
\textbf{Attribute-Based Encryption (ABE)} is a type of public-key encryption scheme where the ciphertext (i.e., an encrypted plaintext) and its corresponding decryption key are linked via attributes~\cite{MAABE,ABE}. In Ciphertext-Policy Attribute-Based Encryption (CP-ABE)~\cite{CP-ABE,MultiAuthorityCP}, a set of such attributes is assigned to potential users. Policies are linked to ciphertexts and articulated as propositional formulae over the attributes. The formulae are evaluated to determine whether a user holds the necessary properties to grant access to the unencrypted data.

\begin{figure}[tb]
	\includegraphics[width=0.9\textwidth]{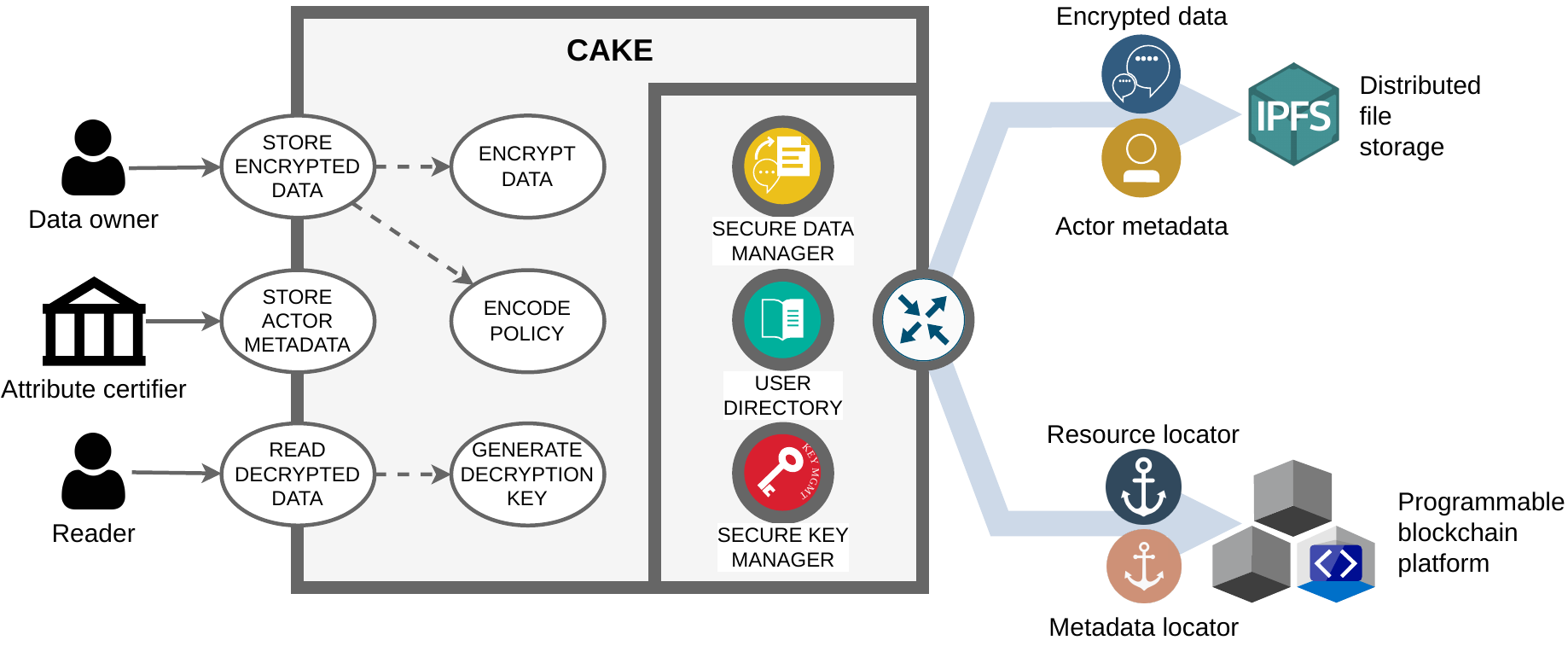}
	\caption{An overview of the CAKE architecture}
	\label{fig:cake:architecture}
\end{figure}
\mysubsubsection{Tool architecture.}
\Cref{fig:cake:architecture} depicts the core traits of CAKE's architecture.
It offers three core functionalities, drawn as use cases in the figure:
\begin{iiilist}
\item storing encrypted data, which in turn requires the encryption of the transmitted artifacts via ABE and the encoding of ciphertext policies to control access;
\item storing actor metadata, mapping ABE attributes to specific users to later determine their suitability to read the stored information;
\item reading decrypted data, which entails the generation of decryption keys depending on the attributes that the requesting user bears.
\end{iiilist}
Those three functionalities are realized by the interplay of three basic components:
\begin{iiilist}
	\item the Secure Data Manager (SDM), which is responsible for the encryption of data based on the policies and the subsequent storage thereof;
	\item the User Directory (UD), recording the association of users with the attributes they bear in the context of the collaborative process enactment;
	\item the Secure Key Manager (SKM), which generates decryption keys for users who wish to read data in clear based on their attributes.
\end{iiilist}
CAKE is interfaced with an IPFS distributed file storage to save the files with encrypted data and with the actor metadata. It resorts to a programmable blockchain platform to record the locators of those files via smart contracts.
The \textit{attribute certifier} writes the actor metadata via UD in a file uploaded onto IFPS, the locator of which is later stored on-chain. 
A \textit{data owner}, namely a process actor that wants to share data with selected users, sends the data in clear and the policies to encrypt it to the SDM. The latter performs the encryption based on the encoded policy, stores the secured data on IPFS, and notarizes the locator thereof on chain. 
To access the data, a \textit{reader} 
asks for a decryption key to the SKM, which in turn retrieves the users' attributes from the UD and uses them to generate the key. 
If those attributes 
satisfy the policy originally used to encrypt the data, the reader can access the contents in clear.
Note that the transactions stored on-chain do not disclose core information. The hash-based resource locator is stored on chain, but the sender of the transaction is the SDM itself, and the recipient is a smart contract. Thus, even if any network node can fetch the public ledger, it cannot extract any information on the exchanged data, its owner, or the intended readers therefrom. 

The detailed explanation of the above passages goes beyond the scope of this demo paper. More information can be found in the paper describing the CAKE approach~\cite{Marangone.etal/BPM2022:CAKE}.
Next, we provide further details about the implementation of CAKE.

%% file: sections/demostrating.tex
\begin{figure}[tb]
	\includegraphics[width=\textwidth]{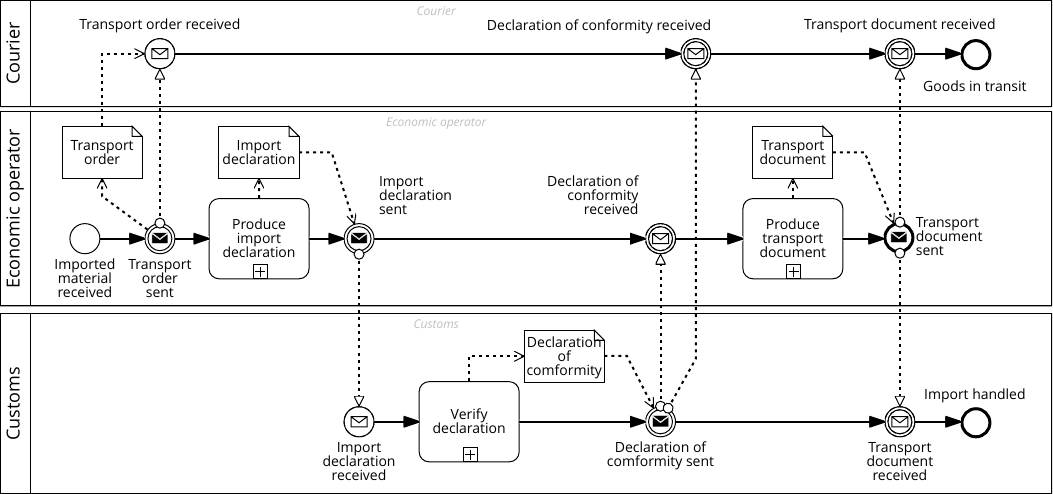}
	\caption{An excerpt of a process workflow in the BRIE project}
	\label{fig:brie:workflow}
\end{figure}
BRIE (Blockchain Register for Import-Export)\textsuperscript{\ref{foot:brie}} is a project aimed at the design and realization of a blockchain-based solution for the monitoring and optimization of international logistics processes. The primary goal is to support stakeholders by facilitating the tracking of shipments, the effective management of pertinent documentation, and the establishment of novel synergies to enhance the management, storage, and transit of goods within Europe.

\Cref{fig:brie:workflow} shows a Business Process Model and Notation (BPMN) collaboration diagram~\cite{Dumas.etal/2018:FundamentalsofBPM} illustrating a model fragment of a process handled in the BRIE project. The process actors involved are the \textit{Courier}, the \textit{Economic Operator}, and \textit{Customs}. 
A new process instance begins when the Economic Operator sends a \textit{transport order} to a Courier. Then, the Economic Operator compiles an \textit{import declaration} for Customs. This document describes the goods, the country of destination, the buyer, and the selected courier for the import. The import declaration is subsequently verified and confirmed by Customs emitting a \textit{declaration of conformity} which can be accessed both by the Economic Operator and the Courier. After this confirmation, the Economic Operator produces a \textit{transport document} with the mode of transport, the Courier, the data and address for goods collection, the expected delivery date, and the delivery address.

Since we utilize ABE, we associate the process actors with users, each having attributes that characterize their role. 
We assume here that the importing country's Chamber of Commerce and the competent ministerial body acted as attribute certifiers to register the actors involved as licensed operators.
In our example, the involved Courier, Economic Operator and Customs are associated with attributes \texttt{courier}, \texttt{economic\_operator}, and \texttt{customs}, respectively.
We represent the participation in the process identified by number {29837} with an attribute recalling the number itself (\texttt{29837}) for short.

\begin{table}[bt]
	\centering
	\resizebox{\textwidth}{!}{%
		\begin{tabular}{|c|c|c|c|}
			\hline
			\textbf{Document}           & \textbf{Sender} & \textbf{Recipients} & \textbf{Policy}                                             \\ \hline
			Transport order    & Economic Operator & Courier    & \texttt{(29837 and ((economic\_operator) or (courier)))} \\ \hline
			Import declaration & Economic Operator & Customs    & \texttt{(29837 and ((economic\_operator) or (customs)))} \\ \hline
			Declaration of conformity &  Customs & Economic Operator; Courier & \texttt{(29837 and ((customs) or (economic\_operator) or (courier)))} \\ \hline
			Transport document        &  Economic Operator & Courier; Customs           & \texttt{(29837 and ((economic\_operator) or (customs) or (courier)))} \\ \hline
		\end{tabular}%
	}
	\caption{Documents exchanged in \cref{fig:brie:workflow} for process instance {29837}}
	\label{tab:documents}
\end{table}
In ABE, policies are linked with ciphertexts and articulated as propositional formulae over attributes. They serve to ascertain whether a user is authorized for access. 
\Cref{tab:documents} contains the encryption policies associated with the aforementioned documents. The one related to the import declaration, e.g., is expressed as \texttt{(29837 and ((economic\_operator) or (customs)))} as it is meant to be accessed by the Economic Operator and Customs involved in case {29837}.
CAKE encrypts every document with the corresponding policy to let only the intended actors read it. 
Notice that if the writer of a document wants to be able to decrypt the shared document later on, they need to include themselves in the set of authorized readers.
Once the document is encrypted, it is uploaded on IPFS, and the resulting resource locator (e.g., \texttt{QmTnDqWf[...]i9wZUgYp}), is stored on the blockchain alongside a unique message ID. 

To access the import declaration later, Customs must ask for a decryption key and obtain the document in clear. 
In this example, the attributes of Customs satisfy the policy used to encrypt the import declaration, so they can obtain the document and read its decrypted content. On the contrary, the Courier has attributes that do not satisfy the policy, so they cannot access that document's content. 
After the Customs agency verifies the conformity of the declaration, the process can progress. As for the transport document, all the three players considered in this example can read the document. Their attributes satisfy the ciphertext policy, so their key is apt for decryption.

The above scenario was used during the final review meeting of the BRIE project and involved a larger integrated platform for logistics data collection and exchange. Next, we provide an overview of the implementation of CAKE and its integration with the BRIE platform.

%% file: sections/implementation.tex
We implemented CAKE and its communication channels in Python. 
The CAKE components expose their interfaces as APIs to a bundled service provider 
to ease communication and integration with other systems.
The communication infrastructure is thus external to the blockchain and IPFS 
and relies on the Secure Sockets Layer (SSL) protocol. We used this protocol to mitigate the risk of packet sniffing by potential malicious third parties aiming to intercept the transmitted data. Moreover, the communication from the data owner to the Secure Data Manager and from the reader to the Secure Key Manager is preceded by an initial authentication phase via a preliminary handshake. Without this security measure, any malicious peer could submit requests on behalf of the authentic reader, having obtained their address and conjecturing a file to which access might be granted.
CAKE allows for the encryption of different types of documents. It is possible to handle a single text file, with the option of applying different policies to different parts thereof. Alternatively, multiple text or binary documents can be uploaded, each being associated with a separate policy. 

The source code of CAKE is openly available at 
\href{https://github.com/apwbs/CAKE}{https://github.com/apwbs/CAKE}.
In the code repository, we provide two implementations of CAKE, distributed within Docker containers: one for the EVM and one for the AVM.
The smart contracts we employ are encoded in Solidity for the EVM
and in PyTeal for the AVM.
They are deployed on the Sepolia testnet%
\footnote{\href{https://sepolia.etherscan.io/}{https://sepolia.etherscan.io/}, accessed 2024-03-11}
and the Algorand testnet,%
\footnote{\href{https://app.dappflow.org/dashboard/home}{https://app.dappflow.org/dashboard/home}, accessed 2024-03-11} respectively.
The AVM-based version of CAKE was used for the BRIE project described in \cref{sec:demostrating}.
To integrate our tool with the BRIE platform, we developed a plug-in named Secrecy and Privacy Enhancer for Ciphered Knowledge (SPECK, available at \href{https://github.com/MichaelPlug/SPECK}{github.com/MichaelPlug/SPECK}),
including a collection of scripts to automatically retrieve information from a shared data repository and interact with the APIs of CAKE. 

Next, we provide a comparative summary of 
research endeavors that relate to CAKE. 

%% file: sections/sota.tex
Numerous research endeavors
have focused on automating collaborative processes utilizing blockchain technology. Weber et al.~\cite{Weber.etal/BPM2016:UntrustedBusinessProcessMonitoringandExecutionUsingBlockchain} introduce a method leveraging this technology to facilitate the conduction of business among parties in the absence of mutual trust. Their work demonstrates how actors can mutually agree on executed behaviors without relying on a central enforcement authority.
López Pintado et al.~\cite{Lopez-Pintado.etal/SPE2019:Caterpillar} introduce Caterpillar, a process execution engine based on Ethereum. Caterpillar enables users to generate process instances and monitor their progress.
Madsen et al.~\cite{Madsen.etal/FAB2018:CollaborationamongAdversaries:DistributedWorkflowExecutiononaBlockchain} investigate the execution of distributed declarative workflows, particularly in situations involving collaboration among adversarial entities. 
Corradini et al.~\cite{Corradini.etal/ACMTMIS2022:EngineeringChoreographyBlockchain} introduce ChorChain, a tool that executes and monitors process choreographies on the Ethereum blockchain platform. 
These studies enhance the fusion of blockchain and process management, unlocking security and traceability opportunities. However, they lack mechanisms to ensure fine-grained access control over data stored on a public platform. In contrast, our work addresses this aspect in a collaborative business process scenario.

Another research area within our domain concerns the privacy and integrity of data stored on-chain. 
Hawk~\cite{Hawk} is a decentralized system that leverages user-defined private smart contracts to execute cryptographic techniques autonomously. In contrast, our approach eliminates the need for custom smart contract encoding, as it relies on on-chain policies for message encryption.
Rahulamathavan et al.~\cite{IoT-ABE} introduce a novel privacy-preserving blockchain architecture tailored for Internet of Things (IoT) applications, utilizing Attribute-Based Encryption (ABE) techniques. While we also utilize ABE, we aim to augment existing software architectures. In contrast, their model alters the blockchain protocol itself.
Benhamouda et al.~\cite{BenhamoudaCanAP} propose a solution enabling a public blockchain to function as a repository for confidential data. In their system, a secret is initially stored on the blockchain, followed by the specification of conditions for its release, with the secret disclosed only if these conditions are satisfied. In contrast, our approach involves the utilization of shared secrets among components. However, it does not entail utilizing the blockchain as a storage system for secret data or disclosing the secret.
Differently from these methodologies, our approach addresses the challenge of regulated data access within a multi-party process scenario. This scenario involves the exchange of multiple information artifacts, where various actors can read specific segments of messages based on access policies.

Wang et al.~\cite{EHR} propose an electronic health record framework integrating Attribute-Based Encryption (ABE), Identity-Based Encryption (IBE), and Identity-Based Signature (IBS) mechanisms with blockchain technology. Unlike the CAKE model, this system design empowers hospitals with patient data ownership while patients delineate access policies. In our architecture, no central authority is intended to manage the data except the data owners, who, in healthcare processes, would be the patients. 
Pournaghi et al.~\cite{MedSBA} propose a framework named MedSBA that leverages blockchain technology and Attribute-Based Encryption. The distinction in their architecture lies in using two private blockchains. Instead, we consider only a public blockchain scenario.

%% file: sections/conclusion.tex
In this paper we presented CAKE,
a tool integrating public blockchain platforms, Attribute-Based Encryption (ABE) and the InterPlanetary File System (IPFS) for controlled data access within multi-party processes. IPFS serves as a tamper-proof repository for storing information artifacts, access policies, and actor metadata. Smart contracts manage user attributes, determine access permissions for process participants, and establish connections to IPFS files for notarization. Thereby, CAKE offers 
the ability to define precise specifications of access privileges, while ensuring data integrity, immutability, non-repudiation, and ultimately facilitating auditability.
The maturity and integration of the tool is testified by its adoption in the context of a real-world cybersecurity project (BRIE)\textsuperscript{\ref{foot:brie}} in the area of international logistics.
Testing the adoption of our tool in further industry settings, thereby gathering feedback, extracting practical implications and devising theory from on-field experience~\cite{Ghaisas,WIERINGA2015136} represents a future, highly interesting endeavor.
Nevertheless, our solution exhibits limitations that we aim to address in future work, too.
To begin with, whenever a data owner wishes to withdraw access to data from a specific reader, the only possibility is modifying the policy and re-encrypting the messages. However, the data previously uploaded on IPFS remains accessible. We are considering InterPlanetary Name System (IPNS) to overcome this limitation. 
Recently, we introduced an alternative approach to blockchain-based secure data sharing in cooperative settings, which divides the tasks of the attribute certification and key forging among multiple computing nodes in a distributed fashion~\cite{Marangone.etal/EDOC2023:MARTSIA}. The full distribution of computing (and the additional overhead it entails) was deemed as unnecessary in the BRIE setting among the stakeholders, due to the involvement of authoritative bodies for the attribution of user metadata and keys in the project. It is in our plans to implement the latter solution on multiple blockchain platforms and reach a level of maturity that is akin to CAKE in order to conduct comparative analyses on the applicability and trade-offs of the two approaches.
Also, we plan to incorporate oracles in our solution to enable smart contracts' validation of off-chain data~\cite{Basile.etal/BPMBCF2021:BlockchainProcessesDecentralizedOracles,DBLP:conf/bpm/MuhlbergerBFCWW20}. This integration empowers system designers to set the balance between complete transparency in the decision-making process and access control. Achieving this equilibrium entails strategically managing the storage of data both on-chain and off-chain, as discussed in~\cite{DBLP:conf/bpm/HaarmannBNW19}.
Finally, combining our solution with techniques for process analytics based on blockchain data~\cite{DBLP:conf/bpm/MuhlbergerBCGL19,Klinkmueller.etal/BCForum2019:ExtractingProcessMiningDatafromBlockchainApplications,DBLP:conf/bpm/HobeckW23} paves the path for future research avenues.